%Paper: hep-th/9508131
%From: gagne@physics.ucla.edu (Darius Gagne)
%Date: Thu, 24 Aug 1995 14:24:11 -0700
%Date (revised): Thu, 21 Sep 1995 17:16:01 -0700

\magnification=1200
\font\stepfour=cmr10 scaled\magstep3
\font\stepone=cmr10 scaled\magstep1

\hfuzz=12pt
\def\makefootline{\baselineskip=36pt\line{\the\footline}}
\nopagenumbers

% slash.tex  --  some slashed characters for use in math mode
%
% "\slashsym<num><sym>" gives slashed version of <sym>; slash is shifted
% to right by <num>mu. Example: "\slashsym9A" looks about right.
\catcode`\*=11  % use * as ordinary char temporarily
\def\slashsym#1#2{\mathpalette{\sl*sh{#1}}{#2}}
\def\sl*sh#1#2#3{\ooalign{\setbox0=\hbox{$#2\not$}
                          $\hfil#2\mkern-24mu\mkern#1mu
                           \raise.15\ht0\box0\hfil$\cr
                          $#2#3$}}
\catcode`\*=12

\def\pslash{{\slashsym8p}}
\def\qslash{{\slashsym7q}}

\def\Aslash{{\slashsym9A}}

\def\dslash{{\slashsym5\partial}}

\rightline{UCLA/95/TEP/22}
\vskip .3in

\centerline{\bf\stepfour Worldline Path Integrals for Fermions}
\medskip
\centerline{\bf\stepfour with}
\medskip
\centerline{{\bf\stepfour Scalar, Pseudoscalar and Vector Couplings}
\footnote{${}^\dagger$} {Research supported in part by National Science
Foundation grant PHY-92-18990}}

\vskip .7in

\centerline{{\stepone Eric D'Hoker and Darius G. Gagn\'e}\footnote{${}^*$}
{E-mail addresses : {\tt dhoker@physics.ucla.edu and
gagne@physics.ucla.edu}}}
\medskip
\centerline{\it Department of Physics and Astronomy}
\medskip
\centerline{\it University of California, Los Angeles}
\medskip
\centerline{\it Los Angeles, CA 90024, USA}

\vskip .8in

\centerline{\bf Abstract}

\vskip .2in

A systematic derivation is given of the worldline path integrals for the
effective action of a multiplet of Dirac fermions  interacting with
general matrix-valued classical background scalar, pseudoscalar, and vector
gauge fields.   The first path integral involves worldline fermions with
antiperiodic boundary  conditions on the worldline loop and generates the real
part of the one loop (Euclidean) effective action.  The second path integral
involves worldline
fermions with periodic boundary conditions and generates the  imaginary part of
the (Euclidean) effective action, i.e. the phase of the fermion functional
determinant.  Here we also introduce a new regularization for the phase of
functional determinants resembling a heat-kernel regularization.  Compared to
the known special cases, our  worldline Lagrangians have
a number of new interaction terms; the validity of some of these terms is
checked in perturbation
theory. In particular, we obtain the leading order contribution (in the heavy
mass expansion)
to the Wess-Zumino-Witten term,
which generates the chiral anomaly.

\vfill
\eject

\footline={\hss\tenrm\folio\hss}

\noindent
{\bf 1. Introduction}

\bigskip

The worldline formulation of quantum field theory (see [1]) provides a powerful
alternative method for the calculation of Feynman diagrams and effective
actions to first order in the loop expansion [2]. When implemented with the
background field method, simplified Feynman rules may be derived and large
numbers of Feynman diagrams may be combined into simple expressions [3]. This
situation is reminiscent of string perturbation theory, where only a single
diagram arises to each order in perturbation theory [4]. Indeed, first
quantized
worldline path integrals closely resemble string perturbation theory and
interestingly, it is in the point particle limit of string theory that its
power
was rediscovered.

Surprisingly however, the  worldline path integral formulation  has not yet
been presented for the quantization of fields with the most general
interactions. While the case of scalar charged particles coupled to Abelian
gauge fields was obtained long ago, the extensions to include charged
fermions (see [2]) or non-Abelian gauge fields has been obtained only recently
[5]. The worldline formulation for a single Dirac fermion in the
presence of a scalar and a pseudoscalar field was obtained in [6] (although
limited to diagrams with an even number of pseudoscalar vertices).

In the present paper, we shall extend the worldline path integral formulation
to
the case of a multiplet of Dirac fermions, coupled to matrix-valued scalar and
pseudoscalar as well as to non-Abelian vector gauge fields, and obtain
explicit formulas for the one loop effective action. The most general case
would include a further axial vector and anti-symmetric tensor field, but this
problem will be treated in a future publication. The methods presented here are
expected to have a suitable extension to the general case.

The starting point of our method is the expression for the fermion effective
action -- coupled to a scalar, pseudoscalar and vector gauge field -- in terms
of the logarithm of a functional determinant of the Dirac operator ${\cal O}$,
suitably continued to Euclidean spacetime. Due to the presence of the
pseudoscalar field, the Euclidean effective action can have both a real and an
imaginary part, respectively even and odd in its dependence on the
pseudoscalar
field. This discussion is presented in Section 2.

The real part of the effective action admits a manifestly global chiral
invariant
and gauge invariant regularized
heat-kernel representation. To show this, we double the number of fermions by
adding their charge conjugates. The real part of the effective action is then
naturally (and apparently uniquely) given as the logarithm of the functional
determinant of
the operator ${\cal O}^\dagger {\cal O}$. Since the operator ${\cal O}^\dagger
{\cal O}$ is now positive, the determinant admits a standard regularized
heat-kernel representation, which is chiral invariant.

Using the coherent states formalism for fermions (and the coordinate-momentum
representation for bosons), we derive a manifestly gauge invariant and globally
chiral invariant worldline path integral representation for the real part of
the
effective action. The corresponding worldline action that enters the path
integral involves worldline fermions obeying {\it antiperiodic boundary
conditions} on the worldline loop and is  even in the Grassmann variables.
(This
is in opposition to [5], where a worldline action was proposed with both even
and odd terms in the Grassmann variables.) The real part of the effective
action
is discussed in Section 3, one of its subtleties is handled in Appendix A and a
comparison with a Feynman diagram is given in Appendix B.

The imaginary part of the effective action is odd in the pseudoscalar field,
proportional to the  Levi-Civita, or totally anti-symmetric
$\varepsilon$-tensor
and may be expressed as the phase of the functional determinant of the Dirac
operator ${\cal O}$. A new regularization by a heat-kernel-like expression of
this phase is presented and is more subtle than the heat-kernel for the real
part of the  effective action.
Several different equally natural choices may be made for the representation,
none of which preserve global chiral symmetry in a manifest way (while gauge
invariance remains manifest), in keeping
with the appearance of the chiral anomaly. One choice is presented in the main
body of the paper in Section 4, while another is discussed in Appendix C.

Using the coherent states formalism, we derive a simple
worldline path integral representation for the imaginary part of the effective
action.  Gauge invariance is manifest while global chiral symmetry is
explicitly
violated as expected. It is shown that the corresponding worldline action
involves worldline
fermions obeying {\it periodic boundary conditions} on the worldline loop and
is even
in Grassmann variables. The zero modes of the periodic worldline fermions
are responsible for producing the Levi-Civita  tensor, which always appears in
these amplitudes. These issues  are
discussed in Section 4, (with an alternative formulation in Appendix C), while
an explicit calculation of the leading order contribution (in the heavy mass
limit) to the
Wess-Zumino-Witten term is given as an example
of
a contribution to the imaginary part of the effective action in Appendix D.

\vskip 0.3in

\noindent
{\bf 2. Effective Action}

\bigskip

	The field theory studied in this paper is that of a multiplet of Dirac
fermions
coupled to a
background scalar $\Phi(x)$, pseudoscalar $\Pi(x)$, and vector gauge field
$A(x)$.  The most general (CPT invariant) classical action for the theory is
given (in Minkowski spacetime) by
$$
S_M[{\bar \Psi} ,\Phi ,\Pi ,A,\Psi ]
=
\int\limits d^4x_M~
{\bar \Psi}^I \bigl [i\gamma_M^{\mu}\partial_{\mu}   - \Phi + i\gamma_M^5\Pi
                                   + \gamma_M^{\mu}A_{\mu}]^{IJ} \Psi^J \; .
\eqno (2.1)
$$
Here, the spacetime metric has signature $\eta = (+~-~-~-)$ and we make use of
the standard conventions
$$
\eqalign{
\{\gamma_M^{\mu},\gamma_M^
{\nu}\} = 2\eta^{\mu\nu} ~,
\qquad
\{\gamma_M^{\mu}, & \gamma_M^5\} = 0 ~,
\qquad
({\gamma_M^5})^2 = 1
\cr
{(\gamma_M^0\gamma_M^{\mu})}^{\dagger} = \gamma_M^0\gamma_M^\mu\;\; ,\;\;
{\gamma_M^5}^{\dagger} = \gamma_M^5\;\; ,\;\; & {\Phi}^{\dagger} = \Phi\;\; ,\;
\; {\Pi}^{\dagger} = \Pi\;\; ,\;\;  A^{\dagger} = A\; .
\cr}
\eqno (2.2)
$$
The superscripts, $I$ and $J$, in the classical action refer to the
internal quantum numbers of the fermion multiplet serving as
reminders that the background fields are matrix-valued.
Coupling constants have been absorbed into the definition of the background
fields for ease of  notation.

Action $S_M$ is sufficiently general to describe, for example, the
gauge coupling of Dirac
fermions in an arbitrary representation ${\cal R}$ (possibly reducible)
of a compact non-Abelian gauge group $G$, as well as the scalar and
pseudoscalar couplings.  Under these $G$-transformations, $S_M$ is gauge
invariant, provided $\Phi$ and $\Pi$ transform under the representation
${\cal R}\otimes {\cal R}^*$ of $G$. The action $S_M$ may also have a global
chiral symmetry
group $H_L\times H_R$ whose generators commute with $G$.  In this case, the
chiral fermions $\psi _L$ and $\psi_R$ transform under some representations
$T_L$ and $T_R$ respectively and $\Phi +i\Pi$ transforms under
$T_L \otimes T_R^*$ by left and right multiplication
respectively. In the special case of QCD-like theories with $N_c$ colors
and $N_f$ flavors for example, we have $G=SU(N_c)$, $H_L = U(N_f)_L$ and
$H_R=U(N_f)_R$. Of course, axial fermion number is in general anomalous with
respect to color at the quantum level. Lastly, notice that the action $S_M$ is
parity conserving.

Also, it is understood that the scalar field
$\Phi$ may assume a vacuum expectation value, rendering the Dirac fermion
massive; in which case it is convenient to single out the mass by decomposing
$\Phi$ as follows
$$
\Phi = m+\varphi
\eqno (2.3)
$$
and where $\varphi$ has zero vacuum expectation value. For $\varphi=0$, we
recover the ordinary Dirac mass term.

The effective action, $W_M[\Phi,\Pi,A]$ for the fermions in the presence of
the background fields may be defined by the functional integral :
$$
e^{iW_M[\Phi ,\Pi ,A]}= \int {\cal D}{\bar \Psi} {\cal D}{\Psi}
                                                    e^{iS_M[{\bar {\Psi}},
                               \Phi ,\Pi ,A,\Psi ]}
\eqno (2.4a)
$$
or by the functional determinant
$$
iW_M[\Phi,\Pi,A] = \log{\rm Det}\, i[i \dslash_M  - \Phi( x)
                  + i\gamma_M^5\Pi( x)
                  + \Aslash_M(x)]\; ,
\eqno (2.4b)
$$
Here, $\Aslash_M$ stands for $ \gamma_M^{\mu}A_{\mu}$.

As it stands, this functional determinant makes sense only perturbatively in
weak fields, e.g. with the help of dimensional regularization. However, it may
be defined also non-perturbatively as a product of eigenvalues after suitable
continuation to Euclidean spacetime. This definition requires an infrared
regularization that renders the spectrum discrete, as well as an ultraviolet
regularization on the infinite number of possibly large eigenvalues. Both
regularizations can
be naturally achieved using heat-kernel methods.

	Analytic continuation is performed by Wick-rotating
$t_M\rightarrow -it_E$ such that
$$
\partial_{t_M} \rightarrow i\partial_{t_E}  \;\;\; , \;\;\;
(p_M)_0 \rightarrow i(p_E)_4\;\;\; {\rm and} \;\;\;
A_0 \rightarrow  iA_4 \; .
\eqno (2.5a)
$$
The gamma  matrices are unaffected
by the continuation.   However, it is useful to change notation and to define
five generators of a Euclidean Clifford algebra, $(\gamma_E)_j \equiv
i(\gamma_M)_j$,
$(\gamma_E)_4 \equiv (\gamma_M)_0$, and $(\gamma_E)_5 \equiv \gamma_M^5$,
satisfying
$$
  \{(\gamma_E)_a,(\gamma_E)_b\} = 2\delta_{ab} \;\; {\rm and} \;\;
  (\gamma_E)_a^{\dagger} = (\gamma_E)_a \;\; {\rm with} \;\; a,b = \mu,5 \; .
\eqno (2.5b)
$$
Thus, under analytic continuation we have also the following changes
$$
\dslash_M \rightarrow i\dslash_E
\qquad
 {\rm and}
\qquad
\Aslash_M \rightarrow i\Aslash_E\; .
\eqno (2.5c)
$$
The hermiticity property of the background fields remains unchanged by the
analytic continuation.  Under the continuation,
the expression (2.4) for the effective action transforms into
Euclidean space as follows (Henceforth, spacetime is taken to be
Euclidean and the subscript $E$ is dropped.):
$$
-W[\Phi , \Pi,A] = \log{\rm Det}\, [{\cal O}] \; ,
\eqno (2.6)
$$
where the operator ${\cal O}$ is defined by
$$
{\cal O} \equiv \pslash  - i\Phi (x) -
                \gamma_5\Pi (x) - \Aslash (x)\; .
\eqno (2.7)
$$
Here, $ p = -i\partial$ is the momentum operator and $x$ should properly be
viewed as the position operator, both of which are hermitian. The  continuation
procedure of the effective action given here, is free of the inconsistencies
mention in  ref.\ [7]:  it was achieved without continuing the classical
action directly -- avoiding the  more delicate Fermi fields.

	As Eq.\ (2.7) involves the position and momentum operator, the
reformulation of the effective action in terms of path integrals has now been
posed as a problem in elementary quantum mechanics.  It will be assumed that
the fields dampen sufficiently fast at infinity so that
spacetime can be effectively compactified. Also, zero modes of ${\cal O}$
-- which would make the effective action blow up --  may always be lifted by
suitable perturbations of the external fields.  Now, the operator ${\cal O}$
has an unbounded spectrum since its signature is
linear in the momentum.  To make use of the heat-kernel regularization, an
operator with a positive real part is required.  Bounded operators may be
obtained most naturally by splitting the effective action into its real and
imaginary parts:
$$
-W_{\Re}[\Phi ,\Pi ,A] - iW_{\Im}[\Phi ,\Pi, A]
                = \ln (\vert{\rm Det}\, [{\cal O}]\vert )
                + i\arg ({\rm Det}\, [ {\cal O}]) \; .
\eqno (2.8)
$$
Both are parity conserving, since the original fermion action is parity
conserving.  In order to understand which diagrams they separately generate,
we use the expansion of the determinant formula (2.6) in weak field
perturbations,
which we shall here give in
dimensionally regularized momentum space (with the vacuum graph deleted)
$$
\eqalign
{
W& ~=
\sum_{n=1}^{\infty}{1\over n}
       \int {d^Dp_1\over (2\pi)^D}\cdots{d^Dp_n\over (2\pi)^D}
 (2\pi)^D\delta^{(D)}(p_1 + \cdots + p_n)\int {d^Dq\over (2\pi)^D}
           ~{\rm tr}{1\over \qslash - im}\cr
&\qquad \times \bigl( i\tilde{\varphi}_1
                    + \gamma_5\tilde{\Pi}_1 + \tilde{\Aslash}_1 \bigr)
                    \cdots
                    {1 \over \qslash - \pslash_1 - \cdots - \pslash_{n-1} -im}
                    \bigl( i\tilde{\varphi}_n + \gamma_5\tilde{\Pi}_n
                    + \tilde{\Aslash}_n \bigr)\; .                       \cr}
\eqno (2.9)
$$
Here, the fourier transforms of the external fields, namely
$\tilde{\varphi}(p)$, $\tilde{\Pi}(p)$ and $\tilde{A}(p)$,  are introduced.
{}From this perturbation expansion, it is easy to see that graphs with an even
number of $\Pi$ legs are real and contribute to $W_{\Re}$ while those with an
odd number of $\Pi$ legs are imaginary and contribute to $W_{\Im}$.
Graphs with an odd number of pseudoscalar legs involve a single $\gamma _5$
and result in contributions proportional to the anti-symmetric tensor
$\varepsilon_{\mu \nu \rho \sigma}$.

\vskip 0.3in

\noindent
{\bf 3. Worldline Path Integral for the Real Part of the Effective
Action}

\bigskip

	The real part of the effective action defined by (2.8),
$$
W_{\Re}[\Phi ,\Pi ,A]
 = -{1\over 2}\ln ({\rm Det}\, [  {\cal O}^\dagger {\cal O}]) \; ,
\eqno (3.1)
$$
automatically accommodates a positive operator, namely
${\cal O}^\dagger {\cal O}$ and so heat-kernel
regularization may be applied here. However, its expression contains terms
linear in the $\gamma$-matrices. Reformulation of this effective action in
terms
of a worldline path integral leads to a worldline action with
terms linear in Grassmann variables. While there is nothing wrong with such
actions from a mathematical point of view [5], they are usually regarded as
physically unacceptable. This difficulty can be handled by doubling the number
of fermions without altering the value of the effective action in the following
way :
$$
W_{\Re}[\Phi,\Pi,A] = -{1\over 2}\ln  {\rm Det}\, [\Sigma] \; ,
\eqno (3.2)
$$
where the operator $\Sigma$ is defined by
$$
{\Sigma} \equiv  \left(\matrix{ 0                      & {\cal O}  \cr
                              {\cal O}^{\dagger} & 0             \cr}
                                                                  \right) \; .
\eqno (3.3)
$$
It is natural to introduce the six $8 \times 8$ gamma matrices [7]
$$
\Gamma_{\mu} \equiv \left(\matrix{ 0            & \gamma_{\mu} \cr
                              \gamma_{\mu} & 0            \cr}\right)\;\; ,\;\;
\Gamma_5     \equiv \left(\matrix{ 0            & \gamma_5     \cr
                              \gamma_5     & 0            \cr}\right)\;\; ,\;\;
\Gamma_6     \equiv \left(\matrix{ 0            & iI           \cr
                              -iI           & 0            \cr}\right)\;\; ,
\;\;
\eqno (3.4a)
$$
which are all hermitian and satisfy
$$
\{\Gamma_A , \Gamma_B\} = 2\delta_{AB}\, {\rm I}_{8 \times 8} \; .
\eqno (3.4b)
$$
In terms of these matrices, $\Sigma$ may now be recast as
$$
\Sigma = \Gamma_{\mu} ({p_\mu} - A_\mu )  - \Gamma_6 \Phi - \Gamma_5\Pi  \; .
\eqno (3.5)
$$
Notice that this operator may be viewed as the six spacetime dimensional Dirac
operator for a gauge field in which the four first components are those of the
gauge field $A_\mu$, while the fifth and sixth components are $\Pi$ and $\Phi$
respectively, and for a Dirac fermion independent of the fifth and sixth
coordinates.

The operator $\Sigma$ still
has an unbounded spectrum but since it is hermitian, its square $\Sigma ^2$ is
positive and the
heat-kernel regularization may be applied to it. As a result, we have
$$
W_{\Re}[\Phi ,\Pi ,A]
 = -{1\over 2}\ln {\rm Det}\, [\Sigma ]
 = -{1\over 4}\ln {\rm Det}\, [\Sigma^2]
 = -{1\over 4}{\rm Tr}\ln [\Sigma^2] \; .
\eqno (3.6)
$$
The trace is over spacetime, spin, as well as internal degrees of freedom.
Now,
the identity, valid for any positive real number ${\cal E}$ and for any
$\sigma$ with $\Re (\sigma)>0$
$$
\ln \sigma= \int_1^\sigma {dt\over t} = -\int_0^{\infty} {dT\over T}
[e^{-{{\cal E}\over 2}T\sigma}-e^{-{{\cal E}\over 2}T}]
\eqno (3.7)
$$
can be duly extended to the case where $\sigma$ is an operator with positive
real eigenvalues, such as $\Sigma^2$.  At last  then, the heat-kernel
regularization is obtained:
$$
W_{\Re}[\Phi ,\Pi ,A] = {1\over 4}\int_0^{\infty}{dT\over T}
                  {\rm Tr}\, e^{-{{\cal E}\over 2}T\Sigma^2}\;\;.
\eqno (3.8)
$$
The constant term has been dropped since its role of subtracting
out ultraviolet divergences will be replaced by dimensional
regularization.  $W_{\Re}$ has been expressed as the trace of the
imaginary-time quantum mechanical evolution operator with Hamiltonian
$\Sigma^2$.  The $T$-integration serves as the sum over all inequivalent
loops [2].  Using $[{p}_{\mu},G(x)]=-i\partial_{\mu}G(x)$ as well as
the anticommutation relations satisfied by the matrices $\Gamma_A$,
the Hamiltonian may easily be worked out to be
$$
\eqalign
{
{\Sigma}^2 &= (p - A)^2 +{i\over 2} \Gamma_{\mu} F_{\mu\nu}\Gamma_{\nu}
	+i\Gamma_{\mu}\Gamma_6 D_\mu \Phi
					+i\Gamma _\mu \Gamma _5 D_\mu \Pi
               			 + \Phi ^2 + \Pi^2
                   + \Gamma_6\Gamma_5 [\Phi,\Pi]\; ,\cr
&\cr
&D_\mu \Phi = \partial _\mu \Phi - i [A_\mu , \Phi] \; ,
\qquad
D_\mu \Pi = \partial _\mu \Pi - i [A_\mu , \Pi] \; ,
\qquad
F_{\mu\nu} = i[D_{\mu},D_{\nu}]\; .
\cr}
\eqno (3.9)
$$
Notice that the operator $\Sigma ^2$ is manifestly covariant under vector gauge
transformations, as well as under global chiral rotations that transform
the Higgs field $\Phi + i \Pi$. Also, this operator is hermitian and positive
by construction and involves $\Gamma$-matrices only to even powers. Thus, it
is suited for the construction of a worldline path integral
representation of the real part of the fermion effective action, which we shall
now derive.

\vskip 0.2in

\noindent {\it Coherent State Formalism}

\medskip

	The standard states of the coordinate-momentum representation
of elementary quantum mechanics,
shall be
used to convert the  bosonic sector of the trace in (3.1) into a bosonic path
integral.  To  handle the fermionic sector analogously, the coherent states
developed in  ref.\ [8] shall be employed.  The key observation here is that
the
matrices $a^+_r$ and $a^-_r$, $r =1,2,3$ defined  by $a^{\pm}_r \equiv {1\over
2}(\Gamma_r \pm i\Gamma_{r+3})$   satisfy standard Fermi-Dirac anticommutation
relations, given by
$$
\{ a^+_r,a^-_s \} = \delta_{rs}
\qquad \qquad
 \{ a^+_r,a^+_s \}
= \{ a^-_r,a^-_s \} = 0 \; .
\eqno (3.10)
$$
Thus, $a^+_r$ and $a^-_r$ are creation and annihilation
operators, respectively, for a Hilbert space with a vacuum defined in the usual
way: $a^-_r\vert 0 \rangle = \langle 0 \vert a^+_r = 0$. To construct coherent
states, we introduce six independent Grassmann variables, $\theta_r$ and
$\bar\theta_r$, $r = 1,2,3$, which anticommute with themselves, and with
the Fermi operators $a_r^+$ and $a_r ^-$. The differentials $d\theta_r$ and
$d\bar\theta_r$ commute with one another and anticommute with all Grassmann
variables and  Fermi operators.  Also, $\theta_r$, $\bar\theta_r$, $d\theta_r$
and $d\bar\theta_r$ commute with the
vacuum $|0\rangle$. The coherent states are defined as follows
$$
\eqalign{
\langle \theta \vert &\equiv
i\langle 0 \vert \prod_{r=1}^3 ( \theta_r - a^-_r )
\;\;\;\;\;\;\;\;\;\;\;
\vert\theta\rangle\equiv\exp \bigl( -\sum_{r=1}^3 \theta_r a^+_r\bigr)
\vert 0 \rangle\cr
\langle \bar \theta \vert &\equiv
\langle 0 \vert \exp \bigl( -\sum_{r=1}^3 a^-_r \bar \theta_r\bigr)
\;\;\;\;\;\;\;\;
\vert \bar \theta \rangle \equiv
i\prod_{r=1}^3 ( \bar\theta_r - a^+_r)\vert 0\rangle\; . \cr}
\eqno (3.11)
$$
The product symbols are understood to order $r$ in the sequence 123.
These states satisfy the defining equations for coherent states
$$
\eqalign{
\langle\theta\vert a^-_r &= \langle\theta\vert \theta_r
\qquad
a^-_r \vert\theta\rangle = \theta_r \vert\theta\rangle
\qquad
\langle\theta\vert\bar\theta\rangle = \exp (\sum_{r=1}^3 \theta_r\bar\theta_r
)\cr
\langle\bar\theta\vert a^+_r &= \langle\bar\theta\vert \bar\theta_r
\qquad
a^+_r \vert\bar\theta\rangle = \bar\theta_r \vert\bar\theta\rangle
\qquad
\langle\bar\theta\vert\theta\rangle
= \exp (\sum_{r=1}^3 \bar\theta_r\theta_r ) \; .
\cr}
\eqno (3.12)
$$
Further useful expressions may be obtained with the help of the Grassmann
integration, defined by
$$
\int\theta_1  d\theta_1= \int\bar \theta_1  d\bar \theta_1 = i \; .
\eqno (3.13)
$$
In particular, we have a {\it completeness relation}
$$
1 = \int\vert\theta\rangle\langle\theta\vert\, d^3\theta
   = \int d^3\bar\theta \,\vert\bar\theta\rangle\langle\bar\theta\vert \;\; ,
 \hbox{with} \;\; , \;\; d^3\theta = d\theta_3 d\theta_2 d\theta_1 \;\; , \;\;
          d^3\bar\theta = d\bar\theta_1 d\bar\theta_2 d\bar\theta_3 \;\; ,
\eqno (3.14)
$$
and it is proven in [8] that traces may be evaluated with these
coherent states as well :
$$
{\rm Tr}(U) = \int d^3\theta \, \langle -\theta\vert U \vert\theta\rangle \; .
\eqno (3.15)
$$
Notice that special care is needed in checking these formulas, since
$\theta _r$, $\bar\theta_r$, $d\theta_r$ and $d\bar\theta_r$ {\it anticommute}
with two of the
coherent states, namely $\langle \theta |$ and $|\bar \theta \rangle$, but
{\it commute} with the other two, namely
$| \theta\rangle$ and $\langle \bar\theta |$.

\vskip 0.2in

\noindent
{\it The Worldline Path Integral}

\medskip

Inserting complete sets of coordinate states and fermionic
coherent states, the trace of
the evolution operator in  Eq.\ (3.8) may be written as
$$
\eqalign{
{\rm Tr}\,    e^{-{{\cal E}\over 2}T{\Sigma}^2}
           &= {\rm Tr_c}\int d^4xd^3\theta \, \langle x,-\theta\vert
              e^{-{{\cal E}\over 2}T{\Sigma}^2}\vert x, \theta\rangle \cr
           &= {\rm Tr_c}\int\limits_{\rm BC}
         \prod_{i=1}^N\Bigl( -d^4x^id^3\theta^i \,
   \langle x^i,\theta^i\vert \exp\bigl[ -{{\cal E}\over 2}{T\over N}
{\Sigma}^2\bigr]     \vert x^{i+1},\theta^{i+1}\rangle\Bigr)\; .\cr}
\eqno (3.16)
$$
Here ${\rm Tr_c}$ denotes the trace only over the internal degrees
of freedom and shall not be explicitly evaluated here. The subscript ${\rm BC}$
denotes the boundary condition $(x^{N+1},\theta^{N+1})=(x^1,-\theta^1)$ on the
$x$ and $\theta$ integrations.  Before evaluating the  matrix
element of the evolution operator, it is useful to evaluate the matrix elements
of the Dirac matrices :
$$
\langle\theta^i\vert\Gamma_A\Gamma_B\vert\theta^{i+1}\rangle
= -\int d^3\bar\theta^{i,i+1}\, \langle\theta^i\vert\bar\theta^{i,i+1}
\rangle\langle\bar\theta^{i,i+1}\vert\theta^{i+1}\rangle\,
2 \; {^i\psi_A}\psi^{i+1}_B \; , \;\;\;\;\;\; A \neq B \; ,
\eqno (3.17)
$$
\noindent where
$$
\eqalign{
\psi^{i+1}_r &\equiv {1\over \sqrt 2}(\theta^{i+1}_r + \bar\theta^{i,i+1}_r)
\;\;\;\; , \;\;\;\;
\psi^{i+1}_{r+3} \equiv {i\over\sqrt 2}(\theta^{i+1}_r - \bar\theta^{i,i+1}_r)
 \cr
{^i\psi_r} &\equiv {1\over\sqrt 2}(\theta^i_r + \bar\theta^{i,i+1}_r)
\;\;\;\;\;\;\;\, , \;\;\,
{^i\psi_{r+3}} \equiv {i\over\sqrt 2}(\theta^i_r - \bar\theta^{i,i+1}_r) \; .
\cr}
\eqno (3.18)
$$
This is easily verified by writing the $\Gamma$-matrices in terms of
the Fermi operators and inserting complete sets of the coherent states
$\vert\bar\theta^{i,i+1}\rangle$ in the appropriate spots. Denoting the
dependency of the Hamiltonian on the momentum operator, background fields and
$\Gamma$-matrices by $\Sigma^2 ( p, \Phi,\Pi,  A, \Gamma_A\Gamma_B )$, the
matrix element of the evolution operator may now be readily computed:
$$
\eqalign{
& \;\;\;\;\; \langle x^i,\theta^i\vert \exp\bigl[ -{{\cal E}\over 2}
{T\over N}{\Sigma}^2\bigr]     \vert x^{i+1},\theta^{i+1}\rangle \cr
&= -\int {d^4p^{i,i+1}d^3\bar\theta^{i,i+1}\over (2\pi)^4}
e^{i(x^i - x^{i+1})\cdot p^{i+1} + (\theta^i - \theta^{i+1})_r\bar
\theta^{i,i+1}_r}\Bigl( 1 - {{\cal E}\over 2}{T\over N}\Sigma^2_i
 + {\rm O}\big[\bigl({T\over N}\bigr)^2\bigr]\Bigr) \; . \cr}
\eqno (3.19)
$$
Here, we have
$$
\Sigma^2_i \equiv \Sigma^2(p^{i,i+1},\Phi^{i,i+1},\Pi^{i,i+1},
A^{i,i+1},2\;{} ^i\psi_A\psi^{i+1}_B )\; ,
\eqno (3.20)
$$
and the fields $\Phi^{i,i+1}$, $\Pi^{i,i+1}$ and $A^{i,i+1}$ denote the
averages of the corresponding fields with superscripts $i$ and $i+1$.
Substituting the matrix element of the evolution operator back into the
expression for the trace and symmetrizing the positions of the Grassmann
variables in the exponentials gives
$$
\eqalignno
{ &{\rm Tr}\, e^{-{{\cal E}\over 2}T{\Sigma}^2}    & (3.21)        \cr
 & = {\rm Tr_c}\int\prod_{i=1}^N \Bigl[{d^4x^id^4p^{i,i+1}d^3
               \theta^id^3\bar\theta^{i,i+1}\over (2\pi )^4}\Bigr]
   \prod_{i=1}^N \Bigl( 1 - {{\cal E}\over 2}
(\tau^i - \tau^{i+1})\Sigma^2_i
 + O\big[{T^2\over N^2}\bigr]\Bigr) \cr
 & \phantom{ = {\rm Tr_c}}
     \times\exp \Bigl( \sum_{i=1}^N [i(x^i-x^{i+1})\cdot p^{i,i+1}
+ {1\over 2}(\theta^i_r - \theta^{i+1}_r)\bar\theta^{i,i+1}_r
- {1\over 2}\theta^i_r (\bar{\theta}^{i-1,i}_r
   - \bar{\theta}^{i,i+1}_r)]\Bigr) \; .&\cr}
$$
Here, the boundary condition for the integrals on $x$, $\theta$ and $\bar
\theta$ is given by
$(x^{N+1},\theta^{N+1},\bar\theta^{0,1})
=(x^1,-\theta^1,-\bar\theta^{N,N+1}) \; .$
The interpolating propertime, $\tau$, has been introduced such that
$\tau^1 = T$, $\tau^{N+1} = 0$ and $\tau^i - \tau^{i+1} = T/N$.
In the limit $N\rightarrow\infty$, the product in the integrand of
(3.21) is simply the standard path ordered exponential.
In this limit, the worldline path integral is obtained:
$$
\eqalignno
{{\rm Tr}\, e^{-{{\cal E}\over 2}T{\Sigma}^2}
 = {\rm Tr_c}\int {\cal D}p\int\limits_{\rm PBC}{\cal D}x & \int
\limits_{\rm APBC}
     {\cal D}\theta{\cal D}\bar\theta
     ~{\cal P}\exp\biggl{\{ }\int_0^Td\tau\Bigl[ i\dot{x}\cdot p
& (3.22) \cr
&
+ {1\over2}\dot\theta _r \bar\theta_r - {1\over 2}\theta_r
     \dot{\bar\theta}_r - {{\cal E}\over 2}
\Sigma^2(p,\Phi,\Pi,A,2\psi_A\psi_B)\Bigr]\biggr{\} } \; ,\cr}
$$
where ${\cal P}$ is the path ordering operator which acts on the
interaction part of the exponential only. The boundary conditions are
periodic (PBC) on $x$, while antiperiodic (APBC) on $\theta $ and
$\bar\theta$:
$(x(T),\theta(T),\bar\theta(T)) = (x(0),-\theta(0), -\bar\theta(0))$.
Also, we have introduced the new variables $\psi_A(\tau)$, defined as
$$
\psi_r(\tau) \equiv {1\over \sqrt 2}[ \theta_r(\tau) + \bar{\theta}_r(\tau)]
\;\;\;\; , \;\;\;\;
\psi_{r+3}(\tau) \equiv {1\over \sqrt 2}i[ \theta_r(\tau)
- \bar{\theta}_r(\tau)]\; .
\eqno (3.23)
$$
To complete this variable change, the fermionic kinetic term is re-expressed in
terms of the $\psi_A$ variables as well
$$
{1\over 2}\dot{\theta}_r\bar\theta_r - {1\over 2}\theta_r\dot{\bar\theta}_r
             = -{1\over 2}(\psi\cdot\dot\psi + \psi_5\dot\psi_5
               + \psi_6\dot\psi_6) = - {1 \over 2} \psi _A \dot \psi _A\; ,
\eqno (3.24)
$$
and the associated boundary condition becomes $(x(T),\psi_A(T) ) =
(x(0),-\psi_A(0))$.  The Jacobian of the transformation is simply absorbed into
the normalization of correlation functions of the $\psi_A$ fields and so
need not be written down explicitly in the path integral expression.

The terms in the path integral which involve the momentum
may be rearranged as follows
$$
i\dot x\cdot{p} - {{\cal E}\over 2}(p - A)^2 = -{{\cal E}\over 2}
(p - A-{i\dot x\over {\cal E}})^2
                                 -{\dot{x}^2\over 2{\cal E}} + i\dot{x}\cdot A
\; .\eqno (3.25)
$$
The term ${i\dot{x}\over {\cal E}}$ may clearly be shifted away by the
$p$-integration.  Showing then that $(p - A)$ may be replaced by $p$ without
changing the value of the path integral is more subtle but proven in
Appendix A.  The following normalization factor is left over
$$
{\cal N} \equiv \int{\cal D}p \, e^{-{{\cal E}\over 2}\int_0^T
                                                  d\tau p^2(\tau)}\; ,
\eqno (3.26)
$$
satisfying
$$
{\cal N}\int\limits_{\rm PBC}{\cal D}xe^{-\int_0^T d\tau
{\dot{x}^2\over 2{\cal E}}} =  (2\pi{\cal E}T)^{-2 } \int d^Dx \; .
$$

The real part of the one-loop effective action can now be recast into its final
form, which is the main result of this Section. It is given by a worldline
path integral with an action that is even in Grassmann fields $\psi$
obeying antiperiodic boundary conditions
$$
W_{\Re}[\Phi,\Pi,A] = {1\over 4}\int_0^{\infty}{dT\over T}
                        {\cal N}\int\limits_{{\rm PBC}}{\cal D}x\int
                        \limits_{\rm APBC}{\cal D}\psi \,
 {\rm Tr_c} \, {\cal P} e^{-\int_0^Td\tau {\cal L}}\; .
\eqno(3.27a)
$$
The worldline Lagrangian, ${\cal L}$, is given by
$$
\eqalign{
{\cal L}(\tau)   &=  {\dot{x}^2\over 2{\cal E}}
            + {1\over 2}\psi_A\dot\psi_A
            - i\dot{x}\cdot A
            + {i\over 2}{\cal E}\psi_{\mu}F_{\mu\nu}\psi_{\nu}
            + {1\over 2}{\cal E}\Phi^2   + {1\over 2}{\cal E}\Pi^2
          \cr &\cr
           & \quad
				+ i{\cal E}\psi_{\mu}\psi_6 D_{\mu}\Phi
            + i{\cal E}\psi_{\mu} \psi_5D_\mu \Pi
            + {\cal E}\psi_6\psi_5[\Phi,\Pi] \; .
\cr}
\eqno (3.27b)
$$
Clearly this path integral preserves manifestly gauge symmetry. While
global chiral symmetry was manifest at the operator level in terms of $\Sigma
^2$ in Eqs. (3.8-9), this symmetry is less clearly seen once the
$\Gamma$-matrices have been represented by a path integral over Grassmann
variables $\psi$. Nonetheless, after Wick-contraction of the fields $\psi$,
global chiral symmetry naturally transpires and we shall still denote it as
manifest.

The worldline path integral of (3.27) reduces to that in [6] in the limit where
the background fields all commute and the gauge field vanishes (their notation
incidentally interchanges $\psi_5 \leftrightarrow \psi_6$).  However, we stress
that from our construction, it is clear that this worldline path integral can
only generate one-loop diagrams with an  even number of pseudoscalar vertices
(which are real valued in Euclidean  space).

	The path integral in (3.27) is a one dimensional quantum field theory.
Since $x(\tau)$ has a constant zero mode, the Green's function of the
corresponding free field operator for the bosonic sector, ${d^2\over d\tau^2}$,
must be defined on the subspace orthogonal to the zero mode:
$$
{1\over {\cal E}}{d^2\over d\tau^2}g_b(\tau - \tau^{\prime})
= \delta (\tau - \tau^{\prime}) - {1\over T}\; ,
\eqno (3.28)
$$
where the solution satisfying $g_b(T - \tau^{\prime})
= g_b(0 - \tau^{\prime})$ is
$$
g_b(\tau - \tau^{\prime}) = -{\cal E}{(\tau - \tau^{\prime})^2\over 2T} +
            {{\cal E} \over 2} \vert\tau - \tau^{\prime}\vert +
\;\;\hbox{constant}
\; .
\eqno (3.29)
$$
Decomposing $x(\tau)$ on the constant zero mode and all the
other modes orthogonal to the zero mode, that is, $x(\tau) = y(\tau) + x$,
where ${d\over d\tau}x = 0$ and $\int_0^T d\tau y(\tau) = 0$, the
fundamental correlation function in the bosonic sector is
$$
\langle\; e^{\int_0^T d\tau y(\tau)\cdot J(\tau)}\;\rangle
= { \int d^Dx \over (2\pi {\cal E}T)^{2-\epsilon }}
\exp \Bigl[ -{1\over 2}
\int_0^Td\tau_1d\tau_2J(\tau_1)g_b(\tau_1 - \tau_2)J(\tau_2)\Bigr]\; ,
\eqno (3.30)
$$
where the dimension of spacetime has been taken to be $D = 4 - 2\epsilon$ in
preparation for dimensional regularization.   For the fermionic sector, the
Green's function is defined as
$$
{d\over d\tau}g_f(\tau - \tau^{\prime}) = \delta(\tau - \tau^{\prime})\; ,
\eqno (3.31)
$$
and must satisfy $g_f(T - \tau^{\prime}) = -g_f(0 - \tau^{\prime})$ due to the
antiperiodicity of $\psi_A$; demanding further that
$g_f(\tau - \tau^{\prime}) = -g_f(\tau^{\prime} - \tau)$ leads to the solution
$$
g_f(\tau - \tau^{\prime}) = {1\over 2}{\rm sign}(\tau - \tau^{\prime})\; .
\eqno (3.32)
$$
The fundamental correlation function in the fermionic sector is
$$
\eqalignno{
& \langle\psi_{A_1}\cdots\psi_{A_N}\rangle &\cr
&= 8\sum_{{\rm all} \;\; j_i = 1}^N\varepsilon_{j_1...j_N}
\delta_{A_{j_1}A_{j_2}}\cdots\delta_{A_{j_{N-1}}A_{j_N}}
{{\rm sign}(\tau_{j_1} - \tau_{j_2})\over 4}\cdots
{{\rm sign}(\tau_{j_{N-1}} - \tau_{j_N})\over 4}\ .&(3.33)\cr}
$$
The normalization
$\langle 1\rangle_{{\rm fermionic}\;{\rm sector}}
= {\rm Tr}({\rm I}_{8\times 8}) = 8$
is to be contrasted with the one found in ref.\ [5] where
$4 \times 4$ gamma matrices are used instead.
\vskip 0.3in

\noindent
{\bf 4. Worldline Path Integral for the Imaginary Part of the
Effective Action}

\bigskip

	In this Section, we present the evaluation of the imaginary part of
the effective action, i.e. the phase of the fermion functional determinant [9].
To this end, we
first derive a new general regulator for the phase of functional determinants
which resembles the heat-kernel regulator in that it involves an evolution
operator and a positive Hamiltonian.
By analogy with the evaluation of the real part of the effective action, we
shall again double the number of fermions so that the heat-kernel-like
representation is even in powers of the $\Gamma$-matrices and so hence
the related worldline action is
even in the Grassmann variables.  First, the doubling of fermions is carried
out as follows
$$
-iW_{\Im}[\Phi,\Pi,A] = i{\rm arg}{\rm Det}[{\cal O}]= {i\over 2}{\rm arg}{\rm
Det}[{\cal O}^2]  = {i\over 2}{\rm arg}{\rm Det}[\Omega ]
\eqno (4.1a)
$$
where we define
$$
\Omega \equiv  \left(\matrix{ 0                      & {\cal O}  \cr
                              {\cal O} & 0             \cr}
                                                                  \right) \; .
\eqno (4.1b)
$$
Using the six  $\Gamma$-matrices defined in Section 3, $\Omega$ and
$\Omega ^\dagger$ (which we shall need shortly as well) may be recast in the
following form :
$$
\eqalign{
\Omega =& ~\Gamma_{\mu}( p_{\mu} - A_\mu ) - \Gamma_5\Pi - \Gamma_7\Gamma_6\Phi
      \cr
\Omega ^\dagger =& ~\Gamma_{\mu}( p_{\mu} - A_\mu ) - \Gamma_5\Pi +
\Gamma_7\Gamma_6\Phi \; ,
\cr}
\eqno (4.2)
$$
where the six dimensional analogue of $\gamma _5$ is defined by
$$
\Gamma_7 \equiv -i\prod_{A=1}^6\Gamma_A
              = \pmatrix{ 1  &  0  \cr
                          0  & -1  \cr}\; .
\eqno (4.3)
$$
As in Section 3, the goal of introducing the $8\times 8$ $\Gamma$-matrices is
to
obtain a final worldline action even in the Grassmann variables. Thus, the
starting point for the imaginary part of the effective action will be the
following expression
$$
-iW_{\Im}[\Phi, \Pi, A] = {1 \over 4} {\rm Tr} \log \Omega -
				{1 \over 4} {\rm Tr} \log \Omega ^\dagger \; .
\eqno (4.4)
$$
At this point, there are different choices for passing from this expression
to a
heat-kernel formulation. These different choices correspond to different
regularizations carried out on the effective action and none of these will
exhibit manifest chiral symmetry, in keeping with the appearance of the chiral
anomaly. In this Section, we shall pursue one particular choice which appears
to
us  most advantageous. An alternative natural choice will be presented in
Appendix C.

We begin by recasting Eq. (4.4) in terms of a single trace with the help of an
auxiliary integration over a parameter $\alpha$ as follows
$$
\eqalign{
-iW_{\Im}[\Phi, \Pi, A]
&
= {1 \over 4} \int _{-1} ^1 d\alpha ~{\partial \over
\partial \alpha} {\rm Tr} \ln \biggl [ {1 \over 2} (\Omega + \Omega ^\dagger)
+ \alpha {1 \over 2} (\Omega - \Omega ^\dagger) \biggr ]
\cr
&
= {1 \over 4} \int _{-1} ^1 d\alpha  ~{\rm Tr} (\Omega - \Omega ^\dagger)
\bigl [  (\Omega + \Omega ^\dagger) + \alpha  (\Omega - \Omega^\dagger) \bigr]
^{-1} \; .
\cr}
\eqno (4.5)
$$
Contributions that are odd in $\alpha$ in the integrand vanish in the integral,
so only the even part may be retained. This allows us to take the average of
the
integrands for $\alpha$ and $-\alpha$, which leads to the following expression
$$
-iW_{\Im}[\Phi, \Pi, A]
= {1 \over 4} \int _{-1} ^1 d\alpha  ~{\rm Tr} \bigl (\Omega ^2 - \Omega
^{\dagger 2} \bigr )
\biggl [  (\Omega + \Omega ^\dagger)^2 + 2 \alpha [\Omega , \Omega ^\dagger] -
\alpha ^2  (\Omega - \Omega^\dagger)^2
\biggr ]^{-1} \; .
\eqno (4.6)
$$
The distinctive advantage of this expression for the imaginary part of the
effective action is that we have produced a formula where the denominator is a
positive operator, as can be seen from the identity
$$
(\Omega + \Omega ^\dagger)^2 + 2 \alpha [\Omega , \Omega ^\dagger] -
\alpha ^2  (\Omega - \Omega^\dagger)^2
=\bigl [  (\Omega + \Omega ^\dagger) + \alpha  (\Omega - \Omega^\dagger) \bigr]
\bigl [  (\Omega + \Omega ^\dagger) + \alpha  (\Omega - \Omega^\dagger)
\bigr]^\dagger \; .
\eqno (4.7)
$$
As a result, we can represent this positive denominator in terms of a
heat-kernel
formula :
$$
-iW_{\Im}[\Phi,\Pi,A]
 = {{\cal E}\over 32}
    \int_{-1}^1 \!\!\! d\alpha ~\int_0^\infty \!\!\! dT \,
    {\rm Tr}\biggl \{
       (\Omega^2 - \Omega^{\dagger 2})
       e^{-{{\cal E}\over 2}T
              \bigl[ {1 \over 4}(\Omega + \Omega^{\dagger})^2 + {1 \over 2}
                     \alpha [\Omega, \Omega ^\dagger]
           - {1 \over 4} \alpha ^2(\Omega - \Omega ^\dagger)^2
\bigr] } \biggr{\} } \; .
\eqno(4.8)
$$
The imaginary part of the effective action, namely the phase of
${\rm Det}[\Omega ]$,  has now been written as
a completely well defined new heat-kernel expression, and we shall now
evaluate the various ingredients in preparation for a path integral
reformulation of it. The insertion works out to be
$$
\Omega^2 - \Omega^{\dagger 2} = 2i\Gamma_7 \omega \; ,
\qquad \qquad
\omega = i\Gamma_5\Gamma_6 \{ \Pi,\Phi \}
           -i \Gamma_{\mu}\Gamma_6 \{ (p_\mu - A_\mu),\Phi \} \; ,
\eqno (4.9)
$$
while the argument of the exponential involves
$$
\eqalign{
{\cal H}_0
&
={1\over 4}(\Omega + \Omega^{\dagger})^2
                   - {1\over 4} \alpha ^2(\Omega - \Omega^{\dagger})^2
\cr
&
= (p-A)^2 +{i\over 2}\Gamma_{\mu}F_ {\mu\nu}\Gamma_{\nu}
              +{\Pi}^2
              + i\Gamma_{\mu}\Gamma_5 D_{\mu}\Pi
               + \alpha ^2{\Phi}^2
\cr
{\cal H} _1
&
=  {1\over 2} \Gamma _7  [\Omega , \Omega ^\dagger ]
= -i \Gamma _6 \Gamma _\mu D_\mu \Phi + \Gamma _5 \Gamma _6 [\Pi, \Phi ] \; .
\cr }
\eqno (4.10)
$$
Putting all these contributions together, we obtain the following heat-kernel
representation
$$
 W_{\Im} [\Phi, \Pi, A]
=
-{{\cal E}\over 16} \int _{-1} ^1 d\alpha \int _0 ^\infty dT~
{\rm Tr} \biggl \{ \Gamma _7 \omega~
e^{-{{\cal E} \over 2} T ({\cal H}_0 + \alpha
\Gamma _7 {\cal H}_1)} \biggr \}\; .
\eqno (4.11)
$$
The expansion of the exponential in powers of $\alpha$ produces non-zero
contributions to the integral only for even powers. Since $\Gamma _7$ always
occurs multiplied by $\alpha$ and commutes with ${\cal H}_0$ and ${\cal H}_1$,
the $\Gamma _7$ in the exponential only occurs to even powers, which just
reduce
to the identity. Thus, the presence of the $\Gamma_7$ factor in the argument of
the exponential is immaterial, and henceforth, we shall omit it. This leads to
the following simplified expression
$$
 W_{\Im} [\Phi, \Pi, A]
=
-{{\cal E}\over 16} \int _{-1} ^1 d\alpha \int _0 ^\infty dT~
{\rm Tr} \biggl \{ \Gamma _7 \omega~
e^{-{{\cal E} \over 2} T ({\cal H}_0 + \alpha
 {\cal H}_1)} \biggr \}\; .
\eqno (4.12)
$$
Notice that in (4.12), the exponential involves terms without
$\Gamma$-matrices and terms quadratic in $\Gamma$-matrices only, exactly as was
the case for the real part of the effective action. Furthermore, the
Hamiltonian ${\cal H} \equiv {\cal H}_0 + \alpha {\cal H}_1$
coincides with the Hamiltonian expression (3.9) of the real case when
$\alpha = 1$.  It is clear from (4.9), (4.10) and (4.12) that the
new regulator is manifestly gauge invariant while global chiral invariance is
not realized in a manifest way due to the presence
of the  $\alpha$-parameter.

	Now, the conversion of (4.12) into a path integral is almost identical
to the case of the real part in Section 3.  The difference is that the
presence of $\Gamma_7$ in the insertion causes the worldline fermions
to have {\it periodic boundary conditions}.  This can be seen by defining
the worldline fermion number operator, ${\cal F}$, in terms of the Fermi
creation and annihilation operators of Section 3:
$$
{\cal F} \equiv \sum_{r=1}^3 {\cal F}_r,\;\;\;\;{\rm with}\;\;\;\;
{\cal F}_r\equiv a_r^+a_r^-
\; .
\eqno (4.13)
$$
Then, $\Gamma_7$ is identical to the fermion number counter $(-1)^{\cal F}$,
also called the ``G-parity operator" [4], and may be  expressed as
$$
(-1)^{\cal F} = \prod_{r=1}^3 (1 - 2{\cal F}_r) = \Gamma_7 \; .
\eqno (4.14)
$$
The presence of $(-1)^{\cal F}$ under a trace changes the boundary conditions
on the worldline fermions, since at the level of coherent states, it operates
as
$$
\langle -\theta \vert (-1)^{\cal F} = i\langle 0 \vert \prod_{r=1}^3
(-\theta_r - a_r^-)(1 - 2a_r^+a_r^-) = i \langle 0 \vert \prod_{r=1}^3
(-\theta_r + a_r^-) = -\langle \theta \vert \; .
\eqno (4.15)
$$
This identity makes it clear that while the trace of (3.15) involves
antiperiodic boundary conditions on the worldline fermions, the presence of
$\Gamma _7$ modifies these into periodic boundary conditions.

	With this preparation, the trace in (4.12) may be readily converted
into a path integral, using (3.22) together with the change to periodic
boundary conditions worked out above. We find
$$
\eqalignno
{
{\rm Tr}\,\Gamma_7\,\omega\, e^{-{{\cal E}\over 2}T{\cal H}}
&
= {\rm Tr_c}\int d^4xd^3\theta \, \langle x,-\theta\vert\, (-1)^{\cal F}\,
     \omega\, e^{-{{\cal E}\over 2}T{\cal H}}\,\vert x, \theta\rangle
& (4.16) \cr
&
= {\rm Tr_c}\int {\cal D}p\int\limits_{\rm PBC}{\cal D}x
     {\cal D}\theta {\cal D}\bar\theta
     \; \omega (p,\Phi,\Pi,A,2\psi_A\psi_B)(\tau = 0)
\cr
 &
\times{\cal P}\exp\biggl{\{ }\int_0^Td\tau\Bigl[ i\dot{x}\cdot p
+ {1\over2}\dot\theta_r\bar\theta_r - {1\over 2}{\theta}_r
     \dot{\bar\theta}_r - {{\cal E}\over 2}
{\cal H}(p,\Phi,\Pi,A,2\psi_A\psi_B)\Bigr]\biggr{\} } \; ,
\cr}
$$
where the periodic boundary conditions (PBC) are
${(x(T),\theta(T),\bar\theta(T)) = (x(0),\theta(0),\bar\theta(0))}$.

Being periodic on the loop, each worldline fermion, $\theta_r$,
$\bar\theta_r$ and $\psi_A$, can be decomposed as the sum of its zero
mode and modes orthogonal to the zero mode (denoted with a prime).  So under
the change of variables from $\theta _r$, $\bar \theta _r$ to $\psi _A$,
there is a Jacobian, $J$, for the zero mode and a Jacobian, $J^{\prime}$,
for the orthogonal modes.  However, $J^{\prime}$ will be absorbed
in the normalization of the correlation functions of the $\psi_A^{\prime}$
so it need not be exhibited explicitly  whereas the zero modes are to be
integrated out and so we must explicitly exhibit $J$.  Thus the measure
${\cal D} \theta {\cal D}\bar \theta \equiv
d\theta_3d\theta_2d\theta_1d\bar\theta_1d\bar\theta_2d\bar\theta_3
{\cal D}\theta^{\prime}{\cal D}\bar\theta^{\prime}$ becomes
$$
{1\over J}d\psi_1d\psi_2d\psi_3d\psi_4d\psi_5d\psi_6{\cal D}\psi^{\prime}
\equiv {1\over J}d^6\psi{\cal D}\psi^{\prime}
= {1\over J}{\cal D}\psi \; ,
\eqno (4.17)
$$
which defines the ordering in the variable change, and so $J$ is easily
worked out as
$$
J = \det \Bigl({\partial \theta,\bar\theta\over \partial\psi}\Bigr) = i \; .
\eqno (4.18)
$$
This identity, ${\cal D}\theta{\cal D}\bar\theta = -i{\cal D}\psi$,
may also be easily checked by direct comparison with the $\Gamma$-matrix
algebra.

As in the previous Section, the terms in the exponential of the path
integral in (4.16) involving the momentum may be arranged as in (3.25)
and again the term
${i{\dot x}\over {\cal E}}$ may be shifted away by the $p$-integration
(which introduces ${i{\dot x}\over {\cal E}}$ into the insertion).  Showing
then that $p - A$ may again be replaced by $p$, even in the insertion, is
proven in Appendix A.  Of course then after this replacement, the term in the
insertion linear in the momentum can not contribute to the path integral.
Hence the path integral of (4.16) acquires the same
normalization factor as in (3.26).
We therefore obtain the final form for the imaginary part of the
effective action, which is the principal result in this Section.
$$
W_{\Im}[\Phi,\Pi,A] = {1\over 8}\int_{-1}^1 d\alpha \int_0^{\infty}{dT}\,
                        {\cal N}\int\limits_{{\rm PBC}}
{\cal D}x {\cal D}\psi
\, {\rm Tr_c} \, {\cal J}(0) \,{\cal P} e^{-\int_0^Td\tau{\cal L}_{\alpha}}
\; .
\eqno(4.19)
$$
The worldline insertion, ${\cal J}(\tau)$, is given by
$$
{\cal J}(\tau) = [2i\psi_{\mu}\psi_6 \dot{x}_{\mu}\Phi
               - {\cal E}\psi_5\psi_6{\{ }\Pi,\Phi {\} }](\tau)
\eqno (4.20)
$$
and the worldline Lagrangian, ${\cal L}_{\alpha}$, is given by
$$
\eqalign{
{\cal L}_{\alpha}&=  {\dot{x}^2\over 2{\cal E}} + {1\over 2}\psi _A\dot\psi _A
            - i\dot{x}\cdot A
            + {i\over 2}{\cal E}\psi_{\mu}F_{\mu\nu}\psi_{\nu}
            + {1\over 2}{\cal E}\alpha ^2\Phi ^2 + {1\over 2}{\cal E}\Pi^2\cr
           & \qquad
            + i \alpha {\cal E}\psi_{\mu}\psi_6 D_\mu \Phi
            + i{\cal E}\psi_{\mu}\psi_5 D_{\mu}\Pi
            + \alpha {\cal E} \psi _5 \psi _6 [\Pi, \Phi ] \; .   \cr}
\eqno (4.21)
$$
The final PBC is  $(x(T),\psi_A(T)) = (x(0),\psi_A(0))$.  Notice
that the terms in this new worldline Lagrangian and insertion bear close
resemblance to the structure of
superstring perturbation theory to one loop order
for odd spin structure [4].  Furthermore, notice that ${\cal L}_{\alpha}$
at ${\alpha} = 1$ is exactly the same worldline Lagrangian used to describe
the real part of the effective action.  The new worldline Lagrangian is again
manifestly gauge invariant but now, due to the explicit presence of the
$\alpha$-parameter, global chiral symmetry is not realized in a manifest way.

	The Green function and perturbative rules for the bosonic sector
of the one dimensional field theory described by (4.19) are the same as
those given for the real case.  In contrast with the real case, the
presence of the zero mode of $\psi_A(\tau)$
requires that the Green function, $G_f$, of the corresponding free field
operator for the  fermionic sector, ${d\over d\tau}$, be defined on the
subspace orthogonal to the zero mode:
$$
{d\over d\tau}G_f(\tau - \tau^{\prime}) = \delta(\tau - \tau^{\prime})
- {1\over T} \;
\eqno (4.22)
$$
and must satisfy $G_f(T - \tau^{\prime}) = G_f(0 - \tau^{\prime})$; demanding
furthermore that $G_f(\tau - \tau^{\prime}) = -G_f(\tau^{\prime} - \tau)$
leads
to the solution
$$
G_f(\tau - \tau^{\prime})
= {1 \over 2} {\rm sign}(\tau - \tau^{\prime}) - {(\tau - \tau^{\prime})\over
T}\; .
\eqno (4.23)
$$
The fundamental correlation function in the fermionic sector is the same as
(3.33) except for the replacements $\psi_{A_j} \rightarrow
\psi^{\prime}_{A_j}$,
${1\over 4}{\rm sign}(\tau_{j_i} - \tau_{j_{i+1}}) \rightarrow {1\over 2}
G_f(\tau_{j_i} - \tau_{j_{i+1}})$ and the normalization
$8 \rightarrow N^{\prime}$.  The new normalization for the
$\psi^{\prime}(\tau)$ path integral, defined as $N^{\prime} \equiv
\int_{\rm PBC}{\cal D}\psi^{\prime}\exp{\bigl[-{1\over 2}\int_0^Td\tau
\psi_A^{\prime}\dot{\psi}_A^{\prime}\bigr]}$, can be easily worked out by
introducing two Grassmann variables, $\eta_1$ and $\eta_2$:
$$
\eqalign{
N &={1\over T}\int d\eta_1 d\eta_2 \int\limits_{\rm PBC}d^6\psi{\cal D}
\psi^{\prime}
\exp\Bigl{\{ }-\int_0^T\bigl[{1\over 2}\psi_A\dot{\psi}_A -
\eta_1 \eta_2 \psi_1(\tau)\cdots
\psi_6(\tau)\bigr]\Bigr{\} } \cr
&= {1\over T}\int d\eta_1 d\eta_2 \, i \, {\rm Tr} \, \Gamma_7
\exp\Bigl({1\over 8}\,\alpha\beta\,\Gamma_1\cdots\Gamma_6\, T\Bigr)= -1\; .
\cr}
\eqno (4.24)
$$
As an application of this worldline path integral formalism for one loop
diagrams with an odd number of pseudoscalar couplings,  we recover the
Wess-Zumino-Witten term in Appendix D.

\vskip 0.3in

\noindent
{\bf 5. Conclusion}

\bigskip

Worldline path integral formulations for the real and imaginary parts of the
one-loop effective action for a multiplet of Dirac fermions coupled to
matrix-valued scalar, pseudoscalar and vector gauge fields have been
systematically derived.

For the real case, this was naturally achieved by a standard heat-kernel
representation of a positive operator, which was then converted into a
worldline
path integral using a simple coherent state formalism for worldline fermions.
We proved that worldline fermions must obey antiperiodic boundary
conditions here.  Gauge invariance and global chiral symmetry are manifest
in this formulation.

For the imaginary case, no simple heat-kernel representation is directly
available. Nonetheless, we obtained a new modified heat-kernel representation
with the help of an additional integration parameter and an insertion operator,
reminiscent of string perturbation theory to one loop for odd spin structure.
We proved that worldline fermions must obey periodic boundary conditions here.
While manifestly
gauge invariant, this formulation does not exhibit chiral symmetry in a
manifest way, and indeed, the Wess-Zumino-Witten term is recovered as an
example
of imaginary contributions to the effective action.

It is now noted that our two path integral formulations could be combined
into a single path integral formulation (for the generation of one-loop
diagrams both even and odd in the number of pseudoscalar couplings)
through a summation over the two spin structures.  This is in accordance with
superstring perturbation theory [4].

The results of ref. [5] and [6] are special cases of our results and are in
agreement within the respective restrictions of the fields. The dimensional
reduction in [6] of a six dimensional gauge coupling to four dimensions is
naturally understood in our reformulation, from the way the fermions are
doubled, both for the case of the real as well as the imaginary part of the
effective action. This doubling was utilized here in order to achieve worldline
actions which are even in the Grassmann variables, showing that the worldline
actions with mixed powers of Grassmann variables, obtained in [5] can naturally
be avoided.

	It appears that the coupling to an axial vector and anti-symmetric
tensor field can be handled in an analogous way to the case treated here.
These cases will be treated in a future publication.

\vskip 0.3in

\noindent {\bf Acknowledgments}

\bigskip

We gratefully acknowledge helpful conversations with Daniel Cangemi on the
coherent state formalism and with Zvi Bern, Christian Schubert and Matthew
Strassler on worldline path integrals. Also, we thank Charles Buchanan and
Sebong Chun for many conversations that directly motivated this research.
E. D. acknowledges the hospitality of the Aspen Center for Physics where part
of
this work was carried out.

\vskip 0.5in

\noindent {\bf Appendix A : Proof of the Momentum Shift}

\bigskip

In this Appendix we prove the momentum shifts used in the path integral
reformulations of the real and imaginary parts of the effective action.

For the real case, it is useful to define
$$
M(\lambda,T)\equiv\int{\cal D}p\;{\cal P}e^{-\int_0^T d\tau [(p - A)^2+
\lambda B]}\; ,
\eqno (A.1)
$$
where $A$ and $B$ are $N \times N$, $\tau$-dependent fields which do not
commute.  We wish to prove that $A$ may be shifted away in the expression for
$M(\lambda,T)$.  To show this, we first diagonalize $A$ by
$A = U^\dagger D U$, where $D$ is a diagonal matrix with entries $D^1,...,D^N$,
and $U^\dagger U = 1$.  Next we note that for $\lambda = 0$ we get
$$
M(0,T) = \lim_{K\rightarrow\infty}\prod_{k=0}^K \Biggl[ U_k^\dagger\,
{\rm Diag}
\biggl( \int_{p_k}\exp \Bigl[ -(p_k - D_k^1)^2 {T\over K}\Bigr],...,
       \int_{p_k}\exp \Bigl[ -(p_k - D_k^N)^2 {T\over K}\Bigr] \biggr) \, U_k
\Biggr] \; .
\eqno (A.2)
$$
Since each diagonal element $D^1,...,D^N$ may be independently shifted away,
we establish that $M(0,T)$ is independent of $A$.  Now we calculate the first
derivative
$$
\eqalignno{
{\partial M\over \partial\lambda}(\lambda,T)
& = -\int_0^T d\tau \int {\cal D}p \; {\cal P}e^{-\int_0^\tau d\tau^\prime
[(p - A)^2+\lambda B]}\ B(\tau)\, {\cal P}e^{-\int_\tau^T d\tau^\prime
[(p - A)^2+\lambda B]}
& (A.3) \cr
& = -\int_0^T d\tau M(\lambda,\tau)B(\tau)M^{-1}(\lambda,\tau)M(\lambda,T)\; .
& (A.4) \cr}
$$
Thus ${\partial M \over \partial\lambda}(0,T)$ is independent of $A$ since
each $M$ in (A.4) is independent of $A$ when $\lambda = 0$.  It is easy to
see that all higher derivatives of $M$ will also be expressible as
linear combinations of products of $B$, $M$ and $M^{-1}$ as in (A.4) so that at
$\lambda = 0$ all higher derivatives of M are also independent of $A$.
Therefore, $A$ may be shifted away in $M(\lambda,T)$.

For the imaginary case, it is useful to define
$$
M^\prime(\lambda,T)\equiv\int{\cal D}p\;(p - A)_0 \; {\cal P}
e^{-\int_0^T d\tau [(p - A)^2 + \lambda B]}\; .
\eqno (A.5)
$$
We shall prove in analogy to the real case that $A$ may also be shifted
away in the expression for $M^\prime(\lambda,T)$ and this then of course
implies that $M^\prime(\lambda,T)$ vanishes.  The analogue of (A.2) is
$$
\eqalign{
&\;\;\;\;\, M^\prime (0,T)\cr
& = U_0^\dagger\, {\rm Diag}
\biggl( \int_{p_0} (p - D^1)_0 \, e^{\bigl[ -(p_0 - D_0^1)^2
{T\over K}\bigr]},...,
\int_{p_0} (p - D^N)_0 \, e^{\bigl[ -(p_0 -D_0^N)^2{T\over K}\bigr]}\biggr)
\, U_0\; M(0,T) \cr
& = \int_{p_0} p_0 \, e^{-{T\over K} p_0^2} \; M(0,T) = 0 \; . \cr}
\eqno (A.6)
$$
Now, the fromula for the first derivative,
${\partial M^\prime \over \partial\lambda}$, is exactly the same as (A.4)
except with $M$ replaced by $M^\prime$.  Thus,
${\partial M^\prime \over \partial\lambda}(0,T) = 0$ since $M^\prime(0,T) =0$.
Similarly, since all higher derivatives are easily expressible as local
products of $B$, $M^\prime$ and $(M^\prime)^{-1}$, we see that all higher
derivatives vanish at $\lambda = 0$.  Therefore $M^\prime (\lambda,T) = 0$.

\vskip 0.3in

\noindent {\bf Appendix B : Comparison with Perturbation Theory, Real Part}

\bigskip

	In this Appendix, the validity of the new term
$\psi_6\psi_5[\Phi,\Pi]$ in the path integral of Eq.\ (3.27) will be checked
by calculating the path integral at order $\Phi^2\Pi^2A^0$ and then comparing
the result with that obtained (at the same order) using ordinary quantum field
theory, namely using Eq.\ (2.9).  This order is in the domain
of applicability of the path integral (having an even number of pseudoscalar
legs) and is the lowest order diagram which
takes into account the noncommuting nature of  $\Phi$ and $\Pi$.  For
simplicity, $\Phi$ is given a vanishing vacuum expectation value (keeping the
fermion massless) and only the pole part of the
graph will be calculated.   Expanding the path integral in Eq.\ (3.27) to order
$\Phi^2\Pi^2A^0$, denoting it by $W_{\Re}[\Phi^2, \Pi^2]$, employing
dimensional regularization and keeping only
those terms which have poles leads to
$$
W_{\Re}[\Phi^2,\Pi^2]
    = {1\over 4} \int\limits_{p_1,...,p_4}\int_0^{\infty}
       {dT\over T}
    (-{{\cal E}\over 2})^2\int_0^T d\tau_1\int_0^{\tau_1}d\tau_2 {\cal X}_{p,T}
    (\tau_1,\tau_2) \; , \eqno (B.1)
$$
where explicitly,
$$
\eqalign
{{\cal X}_{p,T}(\tau_1,\tau_2) &=
 2{\rm Tr_c}(\tilde{\Phi}_1\tilde{\Phi}_3\tilde{\Pi}_2\tilde{\Pi}_4)
    \langle 1\rangle_F
    \langle e^{ix_1\cdot (p_1+p_3)+ix_2\cdot (p_2+p_4)}\rangle_B       \cr
&\; + {\rm Tr_c}([\tilde{\Phi}_1,\tilde{\Pi}_2][\tilde{\Phi}_3,\tilde
    {\Pi}_4])4\langle\psi_{6,1}\psi_{5,1}\psi_{6,2}\psi_{5,2}\rangle_F
    \langle e^{ix_1\cdot (p_1+p_2)+ix_2\cdot (p_3+p_4)}\rangle_B\; .   \cr}
\eqno (B.2)
$$
The background fields
have been written in terms of their fourier transforms:
$$
\Phi(x_i) = \int\limits_{p_i} \tilde{\Phi}_i e^{ip_i\cdot x_i}\;\; ,
{\rm where}\;\; \int\limits_{p_i} = \int {d^Dp_i\over (2\pi)^D} \;\; ,
\tilde{\Phi}_i = \tilde{\Phi}(p_i)\;\; {\rm and} \;\; x_i = x(\tau_i)\; ,
\eqno (B.3)
$$
and likewise for $\Pi$.  The notation $\psi_{A,i} = \psi_A(\tau_i)$
is also used.

	The zero modes in the bosonic correlation functions integrate out as
usual [4, 6] to give overall momentum conservation.  Using (3.30) and
introducing the scaled propertimes
$u_{1,2} \equiv \tau_{1,2}/T$, the $T$-integration of the bosonic correlation
function in either of the two terms in (B.2) becomes
$$
\int_0^{\infty}{dT\over T}T^2(2\pi{\cal E}T)^{\epsilon - 2}
e^{-{{\cal E}\over 2}TK(u_i,p_i)}
= (2\pi{\cal E})^{\epsilon - 2}\Gamma(\epsilon)
[{{\cal E}\over 2}K(u_i,p_i)]^{-\epsilon} = {1\over (2\pi{\cal E})^2}
{1\over \epsilon} + {\cal O}(\epsilon^0)\; ,
\eqno (B.4)
$$
\noindent where $K(u_i,p_i)$ is some function of the scaled propertimes
and the momenta.

	The fermionic correlation function in (B.2) is evaluated using (3.33):
$$
\langle\psi_{6,1}\psi_{5,1}\psi_{6,2}\psi_{5,2}\rangle_F
= -8\cdot 2{{\rm sign}(\tau_1 - \tau_2)\over 4}\cdot
2{{\rm sign}(\tau_1 - \tau_2)\over 4} = -2\; .
\eqno (B.5)
$$

	The integration over the scaled propertimes is trivial when only the
pole part of (B.2) is considered:
$\int_0^1 du_1 \int_0^{u_1} du_2 = {1\over 2!}$.

	Putting everything together, (B.1) becomes
$$
\eqalign{
& \;\;\;\; W_{\Re}[\Phi^2,\Pi^2,{1\over \epsilon}]\cr
& = {1\over 4}{1\over 2!}(-{{\cal E}\over 2})^2
     \int\limits_{p_1,\cdots,p_4}(2\pi)^D
     \delta^{({\rm D})}(p_1 + \cdots + p_4)\cdot
{1\over (2\pi{\cal E})^2}{1\over \epsilon}\cr
& \;\;\;\;\;\times\Bigl{\{ } 2{\rm Tr_c}(\tilde{\Phi}_1\tilde{\Phi}_3\tilde
  {\Pi}_2\tilde{\Pi}_4)\cdot 8
  + {\rm Tr_c}([\tilde{\Phi}_1,
  \tilde{\Pi}_2][\tilde{\Phi}_3,\tilde{\Pi}_4])\cdot 4\cdot (-2)\Bigr{\} }\; .
\cr}
\eqno (B.6)
$$
\noindent Observing that
$$
\eqalign{
& \;\;\;\;\int\limits_{p_1,\cdots,p_4}(2\pi)^D\delta^{({\rm D})}(p_1 + \cdots
  + p_4)
  {\rm Tr_c}([\tilde{\Phi}_1,\tilde{\Pi}_2][\tilde{\Phi}_3,\tilde{\Pi}_4])\cr
& =  \int\limits_{p_1,\cdots,p_4}(2\pi)^D\delta^{({\rm D})}(p_1 + \cdots + p_4)
  [   2{\rm Tr_c}(\tilde{\Phi}_1\tilde{\Pi}_2\tilde{\Phi}_3\tilde{\Pi}_4)
  -   2{\rm Tr_c}(\tilde{\Phi}_1\tilde{\Phi}_3\tilde{\Pi}_2\tilde{\Pi}_4)]
  \; ,\cr}
\eqno (B.7)
$$
\noindent the final result may be written as
$$
\eqalignno{
& \;\;\;\; W_{\Re}[\Phi^2,\Pi^2,{1\over \epsilon}]
     &(B.8)\cr
& = ({1\over 8\pi^2}){1\over \epsilon}\int\limits_{p_1,\cdots,p_4}(2\pi)^D
  \delta^{({\rm D})}(p_1 + \cdots + p_4)[2{\rm Tr_c}(\tilde{\Phi}_1\tilde{\Phi}
     _3\tilde{\Pi}_2\tilde{\Pi}_4) - {\rm Tr_c}(\tilde{\Phi}_1\tilde{\Pi}_2
     \tilde{\Phi}_3\tilde{\Pi}_4)]\; .&\cr}
$$
\noindent Of course, the einbein, ${\cal E}$, canceled out as it should
since it was introduced here as an arbitrary constant.  Furthermore, in the
limit that the fields commute, Eq.\ (B.8) becomes
$$
W_{\Re}[\Phi^2,\Pi^2,{1\over \epsilon}] = ({1\over 8\pi^2}){1\over \epsilon}
\int\limits_{p_1,\cdots,p_4}(2\pi)^D\delta^{({\rm D})}(p_1 + \cdots + p_4)
{\rm Tr_c}(\tilde{\Phi}_1\tilde{\Phi}_3\tilde{\Pi}_2\tilde{\Pi}_4) \; .
\eqno (B.9)
$$
\noindent This last expression agrees with ref.\ [6] modulo a factor of four,
namely, the ${1\over 8}$ here is replaced by ${1\over 2}$ there.  The reason
is simple.  In ref.\ [6], the external fields have been taken to be a
superposition of two plane waves (thus yielding a cross-term factor of 2
for each particle type) while here only a single plane wave is used per
external field.

	 Now, using (2.9) at order $\Phi^2\Pi^2A^0$ gives
$$
\eqalignno{
 W_{\Re}[\Phi^2,\Pi^2]
 = {i^2\over 4} & \int\limits_{p_1,\cdots,p_4}
      (2\pi)^D\delta^{({\rm D})}(p_1 + \cdots + p_4)
      {\rm Tr}[I(\pslash_1,...,\pslash_4)]
& (B.10) \cr
& \times {\rm Tr_c}\bigl[ -(\tilde{\Phi}_1\tilde{\Phi}_3\tilde
      {\Pi}_2\tilde{\Pi}_4 + \hbox{3 cyclic perm.})
+ (\tilde{\Phi}_1\tilde{\Pi}_2\tilde{\Phi}_3\tilde{\Pi}_4
   + \hbox{1 cyclic perm.}) \Bigr] \; . & \cr}
$$
The function $I(\pslash_1,...,\pslash_4)$ is a
standard loop integral whose pole part works out to be
$({1\over 4\pi^2}){1\over \epsilon}$.  Plugging this pole
contribution back in reproduces (B.8).

\vskip 0.3in

\noindent
{\bf Appendix C : Alternative Formulation for the Imaginary Part of
the Effective Action}

\bigskip

In this Appendix, we shall briefly discuss a heat-kernel and worldline path
integral reformulation for the imaginary part of the effective action that is
different from the one developed in Section 4. The reason different
formulations arise for the imaginary part of the effective action is that there
does not seem to be a naturally unique way of obtaining a heat-kernel
formulation for the imaginary part. To us, it seems that the formulation given
in Section 4 is perhaps the most natural one. Yet, the different version
given below may have also certain advantages, while at the same time being less
attractive from other points of view.

The starting point is again relations (4.1) and (4.4). Now, we notice that
$$
{\rm Det} (\Gamma _5 \Omega \Gamma _5)= {\rm Det} (\Omega)
\eqno (C.1)
$$
and thus, we may define a candidate for the imaginary part of the
effective action, which is different from that derived in Section 4, in the
following way
$$
-iW_{\Im}
=   {1 \over 8} {\rm Tr} \ln \{ - \Omega \Gamma _5 \Omega \Gamma _5\}
   - {1 \over 8} {\rm Tr} \ln \{ - \Omega ^\dagger \Gamma _5 \Omega ^\dagger
\Gamma _5\}
\; .
\eqno (C.2)
$$
The operators
$$
\eqalign{
 - \Omega \Gamma _5 \Omega \Gamma _5
&
= H+ \Gamma _7 \omega^{\prime}
\cr
-\Omega ^\dagger \Gamma _5 \Omega ^\dagger \Gamma _5
&
= H- \Gamma _7 \omega^{\prime}
\cr }
\eqno (C.3)
$$
with
$$
\eqalign{
H
&
= (p-A)^2 + {i \over 2} \Gamma _\mu F_{\mu \nu} \Gamma _\nu
+\Phi ^2 - \Pi ^2 - \Gamma _5 \Gamma _\mu \{ (p_\mu - A_\mu ), \Pi \}
\cr
\omega^{\prime}
&
= -i\Gamma _6 \Gamma _\mu D_\mu \Phi - \Gamma _6 \Gamma _5 \{ \Phi, \Pi \}
\cr}
\eqno (C.4)
$$
are  not hermitian, let alone positive. Nonetheless, for
$\Pi = 0$ it is
clear that $H \pm \Gamma_7\omega^{\prime}$ is positive as each may be written
as the product of an operator with its adjoint and therefore, a heat-kernel
representation can be written which is exact in the fields $\Phi$ and $A$
while correct to all orders in a weak field expansion of $\Pi$.

We now have the following expression for the imaginary part of the effective
action in terms of a heat-kernel
$$
\eqalign{
-iW_{\Im}
&
= {1 \over 8} {\rm Tr} \ln (H+\Gamma _7 \omega^{\prime})
            - {1 \over 8} {\rm Tr} \ln (H-\Gamma _7 \omega^{\prime})
\cr
&
= {1 \over 8} \int _{-1} ^1 d \alpha \int _0 ^\infty dT ~{\rm Tr}\biggl \{
\Gamma _7 \omega^{\prime} e^{-T (H+\alpha \Gamma _7 \omega^{\prime})}
\biggr \} \; . \cr}
\eqno (C.5)
$$
We use the same argument as was given in Section 4, to show that the
$\Gamma_7$ factor in the argument of the exponential can be dropped since it is
immaterial. Thus, we are left with the simplified expression
$$
-iW_{\Im}
= {1 \over 8} \int _{-1} ^1 d \alpha \int _0 ^\infty dT ~{\rm Tr}\biggl \{
\Gamma _7 \omega^{\prime} e^{-T (H+\alpha \omega^{\prime})}\biggr \}\; .
\eqno (C.6)
$$
Using the coherent state formalism, we immediately can translate
this expression into a worldline path integral formulation, given by
$$
W_{\Im}
={{\cal E} \over 8} \int _{-1} ^1 d \alpha \int _0 ^\infty dT {\cal N} \int
\limits_{\rm PBC} {\cal D} x {\cal D} \psi ~{\rm Tr _c } \,
{\cal K}(0) ~{\cal P} e^{-\int _0 ^T d \tau {\cal L}'}\; .
\eqno (C.7)
$$
Here, we have the following expressions for the worldline Lagrangian
$$
{\cal L}'   =  {\dot{x}^2\over 2{\cal E}} + {1\over 2}\psi _A\dot\psi _A
            - i\dot{x}\cdot A
            + {i\over 2}{\cal E}\psi_{\mu}F_{\mu\nu}\psi_{\nu}
            + {1\over 2}{\cal E}\Phi ^2
            - {1\over 2}{\cal E}\Pi^2
            + 2 i {\cal E}\psi _\mu \psi _5 \dot x _\mu \Pi + \alpha {\cal E}
              {\cal K}
\eqno (C.8)
$$
and the insertion operator
$$
{\cal K} = -i\psi _6 \psi _\mu D_\mu \Phi + \psi _5 \psi _6 \{ \Pi, \Phi \}\; .
\eqno (C.9)
$$
While this formulation continues to be gauge invariant, it appears asymmetrical
in the way $\Pi$ and $\Phi$ enter and thus chiral symmetry is not manifest.

\vskip 0.3in

\noindent
{\bf Appendix D : Comparison with Perturbation Theory, the
Wess-Zumino-Witten term}

\bigskip

	We make the fermion massive by giving the scalar field a vacuum
expectation value, $\Phi = \varphi + m$.  We shall now
compute the path integral in
(4.19) to order $\varphi^0\Pi^5A^0$ in the background fields.  Only the leading
order in the heavy mass limit is considered.  Expanding the path integral gives
$$
W_{\Im}[\Pi^5] = {1\over 8}\int\limits_{p^1,...,p^5}\int_{-1}^1 d\alpha
\int_0^\infty dT \; (-2{\cal E}m \times {\cal E}^4 \times T^4)
\int_0^1du_1\cdots\int_0^{u_3}du_4 \, {\cal Y}_{p,T}(\alpha ,u_1,...,u_4) \; ,
\eqno(D.1)
$$
where explicitly,
$$
\eqalignno{
{\cal Y}_{p,T}(\alpha,u_1,...,u_4) &= p^1_{\mu_1} \cdots p^4_{\mu_4} {\rm Tr_c}
(\tilde{\Pi}_1\cdots\tilde{\Pi}_5) \langle e^{ix_1\cdot p^1 + \cdots
+ix_4\cdot p^4 + ix_0\cdot p^5} \rangle_B \;e^{-{{\cal E}\over 2}T\alpha^2m^2}
& (D.2) \cr
&\times \int\limits_{\rm PBC} d^6\psi{\cal D}\psi^{\prime}
\psi_{5,0}\psi_{6,0}\psi_{\mu_1,1}\psi_{5,1}\cdots\psi_{\mu_4,4}\psi_{5,4}
\exp\Bigl(-{1\over 2}\int_0^T d\tau \psi_A\dot{\psi}_A\Bigr) \; . & \cr}
$$

	The result of the evaluation of the bosonic correlation function
followed by the $T$-integration is standard.  Performing the new
$\alpha$-integration on this result gives
$$
{1\over (2\pi{\cal E})^2}\Gamma (3)\int_{-1}^1 d\alpha
\Bigl[{{\cal E}\over 2}\alpha^2m^2
+ {{\cal E}\over 2}\kappa (u_i,p^i)\Bigr]^{-3} = -2 \times {1\over 5} \times
{2!\over (2\pi{\cal E})^2}
\times ({2\over {\cal E}m^2})^3 + {\rm O}({1\over m^8}) \; , \eqno (D.3)
$$
where $\kappa (u_i,p^i)$ is some function of the scaled propertimes and the
momenta.

	The fermionic zero mode integration vanishes unless the integrand
contains a factor of $\psi_1\cdots\psi_6$, the integration of which is $+1$.
To obtain such a factor requires that the Lorentz indeces of the momenta must
all be different.  For four different momenta, there are then $4!$
possible momentum combinations which can be expressed succinctly by the
factor $\varepsilon_{\mu_1\mu_2\mu_3\mu_4}p^1_{\mu_1}p^2_{\mu_2}
p^3_{\mu_3}p^4_{\mu_4}$.  The remaining scaled propertime integrations of the
fermionic correlation function becomes
$$
\eqalignno{
\int_0^1du_1\cdots\int_0^{u_3}du&_4\,\,\langle
\psi_{5,2}^\prime\psi_{5,3}^\prime
\psi_{5,4}^\prime\psi_{5,0}^\prime&(D.4)\cr -\psi_{5,1}^\prime\psi_{5,3}^\prime
\psi_{5,4}^\prime\psi_{5,0}^\prime &+ \psi_{5,1}^\prime\psi_{5,2}^\prime
\psi_{5,4}^\prime\psi_{5,0}^\prime - \psi_{5,1}^\prime\psi_{5,2}^\prime
\psi_{5,3}^\prime\psi_{5,0}^\prime + \psi_{5,1}^\prime\psi_{5,2}^\prime
\psi_{5,3}^\prime\psi_{5,4}^\prime \rangle_F \; . & \cr}
$$
Using the appropriate version of (3.33), the first, third and fifth (odd)
correlation functions each easily work out to
be $(1/120 - 1/96)$ while the second and fourth (even)
correlation functions similarly work out to be $(1/ 80 - 1/96)$.
Subtracting the even correlation functions from the odd ones yields
$-{1\over 96}$.

	Putting everything together, (C.1) becomes
$$
W_{\Im}[\Pi^5] = -{1\over 240\pi^2}{1\over m^5}\varepsilon_{\mu_1\mu_2\mu_3
\mu_4}\int\limits_{p^1,...,p^5}(2\pi)^4\delta^{(4)}(p^1 + \cdots + p^5)
p^1_{\mu_1}p^2_{\mu_2}p^3_{\mu_3}p^4_{\mu_4}{\rm Tr_c}(\tilde{\Pi}_1\cdots
\tilde{\Pi}_5) \; .\eqno(D.5)
$$
	Now using (2.9) at order $\varphi^0\Pi^5A^0$ gives (after much algebra)
$$
W_{\Im}[\Pi^5] = -{i\over 5}\int\limits_{p^1,...,p^5} (2\pi)^4\delta^{(4)}
(p^1 + \cdots + p^5)\, (-4im)\, {\rm Tr_c}(\tilde{\Pi}_1\cdots\tilde{\Pi}_5)
\varepsilon_{\mu_1\cdots\mu_4}p^1_{\mu_1}\cdots p^4_{\mu_4}I^{\prime}
(p^i)\; . \eqno (D.6)
$$
The loop integral works out to be ${2!\over (4\pi)^24!}\cdot{1\over m^6} +
{\rm O}({1\over m^8})$.  Plugging the leading order contribution in the
heavy mass limit back in reproduces (D.5).

The contribution to the imaginary part of the effective action
evaluated here is the Wess-Zumino-Witten action [10] to  lowest order in the
fields. The coefficient is found to agree with an explicit Feynman diagram
calculation in the large fermion mass limit, as given also in [11].

\vskip 0.3in

\noindent
{\bf References}

\bigskip

\item{[1]} R.P. Feynman, Phys. Rev. {\bf 80} (1950) 440; R.P. Feynman, Phys.
          Rev. {\bf 84} (1951) 108; J. Schwinger, Phys. Rev. {\bf 82} (1951)
          664; R. Casalbuoni, J. Gomis and G. Longhi, Nouvo Cimento {\bf 24A}
          (1974) 249; R. Casalbuoni, Nuovo Cimento {\bf 33A} (1976) 389;
          L. Brink, P. DiVecchia and P. Howe, Nucl. Phys. {\bf B118} (1977) 76;
          A.P. Balachandran, P. Salomson, B. Skagerstam and
          J. Winnberg, Phys. Rev. {\bf D15} (1977) 2308; R. Casalbuoni, Nucl.
          Phys. {\bf B124} (1977) 93; M.B. Halpern, P. Senjanovic and
          A. Jevicki, Phys. Rev. {\bf D16} (1977) 2476; M.B. Halpern and
          W. Siegel, Phys. Rev. {\bf D16} (1977) 2486; R. Casalbuoni, Nuovo
          Cimento {\bf 64B} (1981) 287.
\item{[2]} A.M. Polyakov, {\it Gauge Fields and Strings} (Harwood Academic
          Pub., 1987).
\item{[3]} Z. Bern and D.A. Kosower, Phys. Rev. {\bf D38} (1988) 1888; Z. Bern
          and D.A. Kosower, Phys. Rev. Lett. {\bf 66} (1991) 1669; Z. Bern and
          D.A. Kosower, Nucl. Phys. {\bf B379} (1992) 451.
\item{[4]} M.B. Green, J.H. Schwarz and E. Witten, {\it Superstring Theory}
          (Cambridge Univ. Press, Cambridge, 1987); E. D'Hoker and D.H. Phong,
          Rev. Mod. Phys. {\bf 60} (1988) 917.
\item{[5]} M.J. Strassler, Nucl. Phys. {\bf B385} (1992) 145.
\item{[6]} M. Mondrag\'on, L. Nellen, M.G. Schmidt and C. Schubert, Phys. Lett.
          {\bf 351B} (1995) 200.
\item{[7]} M.R. Mehta, Phys. Rev. Lett. {\bf 65} (1990) 1983; M.R. Mehta,
          Phys. Lett. {\bf 274B} (1992) 53.
\item{[8]} Y. Ohnuki and T. Kashiwa, Prog. Theor. Phys. {\bf Vol.60} No.2
          (1978) 548.
\item{[9]} D.G.C. McKeon and T.N. Sherry, Ann. Phys. (NY) {\bf 218} (1992) 325.
\item{[10]} J. Wess and B. Zumino, Phys. Lett. {\bf 37B} (1971) 95; E. Witten,
           Nucl. Phys. {\bf B223} (1983) 422.
\item{[11]} E. D'Hoker and E. Farhi, Nucl. Phys. {\bf B248} (1984) 59.

\end